\documentclass[12pt]{article}
\usepackage{epsfig,hyperref,graphicx,amssymb,cite,amsmath,slashed,bbm}
\usepackage[toc,page]{appendix}

\def\beqn{\begin{eqnarray}}
\def\eeqn{\end{eqnarray}}

\def\beq{\begin{equation}}
\def\eeq{\end{equation}}
\def\ba{\beq\new\begin{array}{c}}
\def\ea{\end{array}\eeq}

\newcommand{\ntwo}{${\mathcal N}=2\;$}
\newcommand{\ntwot}{${\mathcal N}= \left(2,2\right)\; $}

\newcommand{\none}{${\mathcal N}=1\;$}
\newcommand{\pt}{\partial}

\newcommand{\qt}{\tilde q}

\begin{document}

\begin{titlepage}

	\begin{flushright}
	FTPI-MINN-18/22, UMN-TH-3806/18\\
\end{flushright}
\begin{center}

{\large \bf Heterotically Deformed Sigma Models on the World Sheet of Semilocal Strings in SQED}

\vspace{5mm}

{\large \bf  E. Ireson$^{\,a}$,  M.~Shifman$^{\,a}$ and \bf A.~Yung$^{\,\,a,b,c}$}
\end{center}

\begin{center}
$^a${\it  William I. Fine Theoretical Physics Institute,
University of Minnesota,
Minneapolis, MN 55455}\\
$^{b}${\it National Research Center ``Kurchatov Institute'', 
Petersburg Nuclear Physics Institute, Gatchina, St. Petersburg
188300, Russia}\\
$^{c}${\it  St. Petersburg State University,
 Universitetskaya nab., St. Petersburg 199034, Russia}
\end{center}




\vspace{5mm}

\begin{center}

\begin{abstract}
A new  two dimensional $\mathcal{N}=(0,2)$  Supersymmetric Non-Linear Sigma Model  describes the dynamics
of internal moduli of  the BPS semi-local vortex string supported in four dimensional $\mathcal{N}=2$ SQED.  While the core of these strings is very similar to Abrikosov-Nielsen-Olesen vortices, they are defined with a characteristic size modulus, much like the instanton lump size. This entails that the constituting fields of the vortex do not decay exponentially, as one goes far away from the core of the string, but as a rational function. The appearance of an extra scale in the problem also allows for an explicit, analytic, approximate solution to be written for the BPS equation, surprisingly. 

Despite the conceptually large differences between semi-local and non-Abelian vortices, it appears that the moduli structures have one main common feature, both undergo the same kind of heterotic deformation when a supersymmetry breaking potential term is added to the spacetime theory, moving from $\mathcal{N}=2$ to $\mathcal{N}=1$. 

 By adding a mass term for the gauge scalar multiplet, a heterotic deformation develops on the worldsheet, which breaks supersymmetry down to $(0,2)$ by coupling supertranslational fermionic zero modes to supersize ones. Such an interaction between zero modes of two different sectors was already hypothesized and subsequently found for non-Abelian strings, providing a neat way of circumventing accidental supersymmetry enhancement via Zumino's theorem. We find that, for small values of the spacetime mass term, an entirely analogous term develops on the worldsheet of semi-local strings.
\end{abstract}
\end{center}

\end{titlepage}

\section{Introduction}
\setcounter{equation}{0}

Vortices with non-Abelian gauge groups (usually $U(N_c)$), as well as extended flavour symmetry, are host to a wealth of unique and surprising properties (\cite{Hanany:2003hp},\cite{Auzzi:2003fs},\cite{SYmon},\cite{Hanany:2004ea},\cite{Gorsky:2004ad},\cite{Eto:2006mz} and \cite{Hindmarsh:1992yy},\cite{Achucarro:1999it},\cite{Shifman:2006kd},\cite{Eto:2007yv}). Non-Abelian colour symmetry leads to non-Abelian strings, which bear a more complex charge structure than in the Abelian Higgs model. This is materialised by an internal  degree of freedom, an undetermined modulus that points in a certain direction in an internal symmetry space, found to be $\mathbb{CP}(N_c-1)$. 

As a consequence, quantising the soliton leads to the study of fluctuations of these parameters in time and along the length of the string, i.e. a two dimensional non-linear sigma model which captures the physics of the vortex string worldsheet. Much is known about the maximally supersymmetric NLSM. When  considering a lesser number of supercharges, one finds that the worldsheet theory becomes a particular type of heterotically deformed, $(0,2)$ supersymmetric Non-Linear Sigma Model. Indeed, it is possible to construct non-Abelian vortices from spacetime field theories with fewer supersymmetries than $\mathcal{N}=2$, for instance by adding a mass term to the  scalar multiplet components of the full gauge supermultiplet, making the spacetime theory $\mathcal{N}=1$, and then to observe the consequences on the worldsheet.

It was originally suggested by Shifman and Yung \cite{Shifman:2005st} that the resulting NLSM would have at least $\mathcal{N}=(1,1)$ supersymmetry, with extra fermionic degrees of freedom. In addition, this process does not spoil the K\"{a}hler nature of the target space at hand, thus would lead to an enhancement back to the full $(2,2)$ theory. This statement often goes by the name of Zumino's theorem \cite{Zumino:1979et}. It did not seem very surprising that these objects benefited from supersymmetric enhancement, since it had been previously proven that this exact phenomenon happens on domain walls \cite{Ritz:2004mp}. 

This came into tension with a different perspective offered by Edalati and Tong \cite{Edalati:2007vk}, who, with the help of a brane model, suggested that this statement was untrue-- while $\mathbb{CP}(N_c-1)$ alone can indeed not be deformed in a way that breaks some but not all of the supersymmetry, the \textit{full} target space that the string explores is $\mathbb{C}\times\mathbb{CP}(N_c-1)$. Indeed, in addition to the internal gauge modulus, there is an ever-present translational modulus which describes the position of the string in the transverse directions. This degree of freedom, and its supersymmetric partners, are usually completely decoupled from whatever internal structure the string may also have. Edalati and Tong argued that, from the worldsheet perspective, it is possible to construct a term that mixes the fermionic sectors in both components of this target manifold (the super-translational and super-orientational fermions) in fully target space invariant way, without entailing a deformation of the manifold itself, thus producing an $\mathcal{N}=(0,2)$ theory.\footnote{For a discussion of general aspects of 2D $\mathcal{N}=(0,2)$ theories see e.g. \cite{Witten:2005px}.}

This hypothesis was then proven explicitly when this term was derived from the ground up in the spacetime theory \cite{Shifman:2008wv}. It was indeed the case that fermionic zero modes in different sectors have some overlap and do not decouple when the supersymmetry breaking potential is turned on, producing exactly the Edalati-Tong heterotic deformation. Many properties of the worldsheet theory were then investigated (\cite{Cui:2010si},\cite{Cui:2011rz}\cite{Chen:2014efa}).

It is therefore relevant to observe if this phenomenon happens for the other type of internal modulus that a generic vortex string may possess: the size modulus. When the number of flavours $N_f$ exceeds the number of colours, the BPS vortex string that occurs in such a theory is no longer fully local. That is, while in a usual Abrikosov-Nielsen-Olesen string every field that constitutes the vortex decays exponentially at a certain distance away from the core, it is found that the fields in a flavour-enhanced string decay as rational functions, defined by a characteristic arbitrary size modulus \cite{Achucarro:1999it}, in a very analogous fashion to the size parameter of the instanton solution\cite{Belavin:1975fg}. Rather surprisingly, the appearance of this extra scale, provided it is much larger than the core width, allows an explicit analytic solution to the BPS equations, albeit an approximate one, to be written.

Such semi-local strings also present idiosyncratic challenges to investigate: because its constitutive fields decay so slowly, the theory requires an infra-red cutoff mechanism in order for integration over the directions transverse to the string to regulate it: such integrals are borderline divergent, logarithmically. However, with this compromise alone, it is then possible to create a consistent worldsheet picture of the string. It has been argued that this was no obstruction to further analysis, as any large logarithmic factor can simply be removed by wavefunction normalisation, so that we should expect the worldsheet picture to make sense in any case\cite{Shifman:2011xc},\cite{Koroteev:2011rb}. This led to some very fruitful investigation on the dynamics of these semi-local strings:  most recently, it was found that a non-Abelian  semi-local vortex string, with two colours and four flavours, is conformal and has a full 10D target space and is therefore a true critical superstring \cite{SYcstring},\cite{Koroteev:2016zqu}.

In this work, we wish to start by investigating the possibility of such heterotically deformed worldsheets in the simplest field theory that bears these semi-local vortices, namely $\mathcal{N}=2$ SQED with two flavours. Even in this simple setup there is a wealth of unique phenomena that have become apparent: it was recently found that these basic semi-local vortices, once made closed, can have an extra type of internal winding number, in addition to the usual vortex number, and that both of them would combine to form a soliton with non-zero Hopf index \cite{Ireson:2018bdw}.

 After checking some of the basic building blocks of the worldsheet theory, we turn on a 4D mass deformation $\mu$, and attempt to solve the modified Dirac equations for the fermion zero modes. At small $\mu$ the picture is very clear, these zero modes become non-holomorphic (in a precise sense to be explained in time), thus allowing for a non-zero overlap between supertranslational and supersize modes of the expected shape:
 \begin{equation}
 	\zeta_R  \partial_L\bar{\rho} \chi_R + \text{ H.c.}
 \end{equation}
 This is formally identical to the kind of term derived in the non-Abelian string case, being naturally constrained by target-space geometry.


\section{Bulk theory}
\setcounter{equation}{0}

Our basic four dimensional model is a   \ntwo supersymmetric Abelian $U(1)$  gauge theory deformed by a \none mass term
$\mu$ for the neutral  gauge scalar supermultiplet, in the following way.
 The \ntwo vector multiplet contains  the gauge bosons $A_{\mu}$, two gauginos $\lambda^{\alpha 1}$ and 
$\lambda^{\alpha 2}$ and the complex neutral scalar field $a$, where $\alpha$ is the spinor index, $\alpha=1,2$. The complex scalar $a$ and one of the gauginos $\lambda^2$ form a   neutral \none chiral supermultiplet 
${\mathcal A}$. Adding a mass $\mu$ to this neutral supermultiplet breaks 
\ntwo supersymmetry in the bulk down to \none. In the limit of $\mu \to \infty$ the neutral multiplet decouples and the theory flows to \none SQED.

The model also has the matter sector consisting of $N_f=2$ ``electron" matter hypermultiplets charged with respect to the gauge $U(1)$ . 
In addition, we will introduce a  Fayet--Iliopoulos $D$-term for the $U(1)$ gauge field
which triggers the scalar electron condensation.

Let us first discuss the undeformed theory with
 \ntwo$\!.$ The  superpotential has the form
 \beq
{\mathcal W}_{{\mathcal N}=2} =\frac{1}{\sqrt 2} \sum_{A=1}^2
 \tilde Q_A {\mathcal A}
Q^A \,,
\label{superpot}
\eeq
where $Q^A$ and $\tilde Q_A$ ($A=1,2$) represent two
matter  hypermultiplets. The flavor index is denoted by $A$.  

Next, we add a superpotential,
\beq
{\mathcal W}_{br}=\frac{\mu}{2} {\mathcal A}^2,
\label{msuperpotbr}
\eeq
 Clearly, the mass term (\ref{msuperpotbr}) splits \ntwo supermultiplets, breaking
\ntwo supersymmetry down to \none.

Note that in \eqref{superpot} we set the electron masses to zero.
As was shown in \cite{Shifman:2005st} and \cite{Edalati:2007vk} (see also the review 
 \cite{SYrev}), in this case the deformed theory supports 1/2 BPS -saturated
flux-tube solutions at the classical level. The massive versions of the deformed \ntwo theory
were studied in \cite{IYmu,IYmusemi}

The bosonic part of our $U(1)$  theory has the form
\beq
S=\int d^4x \left\{
\frac1{4g^2}\left(F_{\mu\nu}\right)^2
+ \frac1{g^2} \left|\partial_{\mu}a\right|^2 
 +\left|\nabla_{\mu}q^{A}\right|^2 + \left|\nabla_{\mu} \bar{\tilde{q}}^{A}\right|^2
+V(q^A,\tilde{q}_A,a)\right\}\,,
\label{model}
\eeq
where
\beq
\nabla_\mu=\partial_\mu -\frac{i}{2}\; A_{\mu},
\label{defnabla}
\eeq
while $g$ is the gauge  coupling constant. Note, that we work in the Euclidean space.

The potential $V(q^A,\tilde{q}_A,a)$ in the Lagrangian (\ref{model})
is a sum of  various $D$ and  $F$  terms,
\beqn
V(q^A,\tilde{q}_A,a) &=&
\frac{g^2}{8}
\left(\bar{q}_A q^A - \tilde{q}_A \bar{\tilde{q}}^A-\xi\right)^2
+\frac{g^2}{2}\left| \tilde{q}_A q^A +
\sqrt{2}\, \mu a
 \right|^2
\nonumber\\[3mm]
&+&
\frac12\sum_{A=1}^2 |a|^2 \left[ |q^A|^2 + |\bar{\tilde{q}}^A|^2\right]
,
\label{pot}
\eeqn
where the sum over repeated flavor indices $A$ is implied.
We also introduced the Fayet--Iliopoulos  $D$-term for the $U(1)$  field,
with the FI parameter $\xi$ in (\ref{pot}).
Note, that the Fayet--Iliopoulos term does not
break \ntwo supersymmetry \cite{matt,VY}. The parameter which does
break  \ntwo   down to \none is $\mu$ in (\ref{msuperpotbr}). 

Let us review briefly the vacuum structure and the mass spectrum of perturbative excitations
in our bulk model \eqref{model}, see \cite{SYrev} for details.

The Fayet--Iliopoulos term triggers the spontaneous breaking
of the gauge symmetry. The vacuum expectation values (VEV's)
of the scalar electrons (selectrons) can be chosen as
\beq
\langle q^{A}\rangle =\sqrt{
\xi}\, \left(
\begin{array}{c}
1  \\
0 \\
\end{array}
\right),\,\,\,\langle \bar{\tilde{q}}^{A}\rangle =0,
\qquad
 A =1,2\,,
\label{qvev}
\eeq
while the VEV of the neutral scalar field vanish,
\beq
\langle a\rangle =0.
\label{avev}
\eeq

The choice of vacuum in \eqref{qvev} is not unique, our theory has a Higgs branch, a manifold in the space of VEV's of $q^A, \qt_A$ fields where  the scalar potential \eqref{pot} vanish. The dimension of this non-compact Higgs branch
is four. To see this, note that we have eight real scalars $q^A, \qt_A$ subject to three
conditions associated with vanishing of two terms in the first line in \eqref{pot}. Also one phase is gauged. Overall we have 
\beq
{\rm dim} {\mathcal  H} = 8-3-1=4,
\label{dimH}
\eeq 
 which is the dimension of the Higgs branch. Four massless scalars correspond
to the lowest components of one short hypermultiplet. 

A generic vacuum on this Higgs branch does not support BPS
string solutions. The reason is that for a generic vacuum the mass of the photon is not equal to the mass of the Higgs field, the condition needed for a string to be BPS.
However, the compact two dimensional base of the Higgs defined by the condition
\beq
\langle \tilde{q}_{A}\rangle =0 
\eeq
does support BPS strings \cite{VY,EvlY}. Below in this paper we restrict ourselves  to the base of the Higgs
branch and since all vacua on the base are physically equivalent we take the vacuum \eqref{qvev} as a particular representative.

Since the  $U(1)$  gauge group is broken by selectron condensation,
the  gauge boson becomes massive. From (\ref{model}) we get the photon mass
\beq
m_{\gamma}=\frac{g}{\sqrt{2}} \sqrt{\xi}\,,
\label{phmass}
\eeq

To get the masses of the scalar bosons we expand the potential (\ref{pot})
near the vacuum (\ref{qvev}), (\ref{avev}) and diagonalize the
corresponding mass matrix. Then, one component of the eight real scalars $q^A, \qt_A$, namely   ${\rm Im}\, q^1$ 
is eaten by the Higgs mechanism.  Another  component, namely ${\rm Re}\, q^1$
 acquires  a mass (\ref{phmass}), equal to the mass of the photon. It becomes
a scalar component of  the  massive \none vector $U(1)$  gauge multiplet. This component is the Higgs field in our theory, since it develops VEV, see \eqref{qvev}. The coincidence of masses ensures presence of BPS strings in our vacuum.

Other four real scalar components of the fields $\tilde{q}_{1}$  and $a$
produce the following states: two states acquire mass
\beq
m^{+}=\frac{g}{\sqrt{2}}  \sqrt{\xi\lambda^{+}}\,,
\label{u1m1}
\eeq
while the mass of other two states is given by 
\beq
m^{-}=\frac{g}{\sqrt{2}}\sqrt{\xi\lambda^{-}}\,,
\label{u1m2}
\eeq
where $\lambda^{\pm}$ are two roots of the quadratic equation
\beq
\lambda^2-\lambda(2+\omega^2) +1=0\,.
\label{queq}
\eeq
 Here we introduced  \ntwo supersymmetry breaking
parameter, 
\beq
\omega=\frac{g^2\mu}{m_{\gamma}}\,.
\label{omega}
\eeq

In the large-$\mu$ limit
the larger mass $m^{+}$ becomes
\beq
m^{+}= m_{\gamma} \omega=g^2\mu\,.
\label{amass}
\eeq
Clearly, in the limit $\mu\to \infty$ this is 
the mass of the heavy neutral
scalar $a$. At $\omega\gg 1$ this field decouple and
can be integrated out.

In this limit the scalar $\qt_1$ becomes a lowest component of the chiral multiplet with
the lower mass $m^{-}$. Equation (\ref{queq}) gives for this mass
\beq
m^{-}= \frac{m_{\gamma}}{\omega}
= \frac{\xi}{2\mu}\,. 
\label{light}
\eeq

Furthermore, the four real components $q^2, \qt_2$ of the second flavor are massless and live on the Higgs branch.
In  the limit of infinite $\mu$  mass \eqref{light} tend to zero.
This fact reflects the enhancement of the  Higgs branch in \none SQED.

Below we will also need the 
  fermionic part of the action  of  the model (\ref{model}),
\beqn
S_{\rm ferm}
&=&
\int d^4 x\left\{
\frac{i}{g^2}\bar{\lambda}_f \bar{\pt}\hspace{-0.65em}/\lambda^{f}
+ \bar{\psi}_A i\bar\nabla\hspace{-0.65em}/ \psi^A
+ \tilde{\psi}_A i\nabla\hspace{-0.65em}/ \bar{\tilde{\psi}}^A
\right.
\nonumber\\[3mm]
&+&
\frac{i}{\sqrt{2}}\,\left[ \bar{q}_{Af}(\lambda^f\psi^A)+
(\tilde{\psi}_A\lambda_f)q^{fA} +(\bar{\psi}_A\bar{\lambda}_f)q^{fA}+
\bar{q}^f_A(\bar{\lambda}_f\bar{\tilde{\psi}}^A)\right]
\nonumber\\[3mm]
&+&
\left.
\frac{i}{\sqrt{2}}\, a (\tilde{\psi}_A\psi^A)
+\frac{i}{\sqrt{2}}\, a (\bar{\psi}_A
\bar{\tilde{\psi}}^A)
 -\frac{\mu}2 (\lambda^2)^2
\right\}\,,
\label{fermact}
\eeqn
where 
 $(\psi^{\alpha})^{A}$ and $(\tilde{\psi}^{\alpha})_{A}$ are matter fermions.
Contraction of the spinor indices is assumed
inside  parentheses.
We write the selectron fields in (\ref{fermact}) as doublets of the $SU(2)_{R}$
group which is present in \ntwo theory
\begin{equation}
q^{fA}=(q^A,\bar{\tilde{q}}^A)\,,
\end{equation}
where $f=1,2$ is the $SU(2)_R$ index, this makes manifest the existence of two sets of supersymmetry operators in the \ntwo case.
Similarly,
$\lambda^{\alpha f}$  stands for
the gaugino $SU(2)_R$ doublet.
Note that the last   term is the  \none deformation
in the fermion sector of the theory induced by the breaking parameter $\mu$.
It involves only $f=2$ component of $\lambda$ explicitly breaking
the $SU(2)_{R}$ invariance.

From \eqref{fermact} one can see that fermions of the second flavor in much the same way as  bosons are massless in the vacuum \eqref{qvev}. This will be important later.

\section{Semilocal strings in the $\mathcal{N}=2$ theory}
\setcounter{equation}{0}

\subsection{Vortex BPS Equations for a Static Solution}

We work in Euclidean space, labelling our coordinates $(t,x,y,z)$. We will assume that the string we produce is aligned in the $z$ direction. 


 As we explained in the previous section the potential \eqref{pot}  has an infinite Higgs branch and we restrict ourselves to its base submanifold with $\tilde{q}^{1,2}=0$. The base of the Higgs branch is then now compact and defined by 
 \begin{equation}
 	|q^1|^2 + |q^2|^2 =\xi
 \end{equation}
 where both $q^{1,2}$ are complex fields, so  the base of the Higgs branch has the structure of $\mathbb{CP}(1)$. At spatial infinity, the vortex configuration is expected to wrap around the vacuum manifold in a non-trivial way: thus we expect that the vortex will behave like the $\mathbb{CP}(1)$ instanton lump solution at large distances from the core, while close to the core it should behave just like a standard ANO string. The instanton lump is endowed with a dimensionful modulus, a size parameter $\rho$ which controls the spreading of the solution in space\footnote{for details see e.g. \cite{Rajaraman:1982is}}: the vortex should be similarly spread out away from the core, this is why it is called semi-local.
 
 Let us introduce a number of profiles for the various bosonic fields in the theory:
\begin{align}
	q^{1A}&\equiv q^{A}=\left(\begin{array}{cc}
	\phi_1(r)  \\ 
	\phi_2(r) e^{-i\theta }
	\end{array}  \right),\quad  q^{2A}\equiv -i\tilde{q}^{A} =0\nonumber \\ A_i&=\varepsilon_{ij}\frac{x^j}{r^2}f(r)
\end{align}
Here we assume boundary conditions
\beq
\phi_A(0) =0, \quad \phi_1(\infty)=\sqrt{\xi}, \quad \phi_2(\infty)= 0, \quad f(0) =1, \quad f(\infty ) =0
\label{bc}
\eeq
which ensure that the scalar fields tend at $r\to\infty$ to their vacuum expectation values \eqref{qvev}.
We have defined this Ansatz in the singular gauge, where $A$ will be ill-defined at $0$ but decay at infinity. We will assume that all of the profile functions are positive in order to fix various sign choices related to supercharges.

From the supersymmetry transformations of our initial theory, we obtain BPS equations and also we define which fermionic variations are preserved by our choices, so as to preserve $\epsilon^{12}Q_{12},\,\epsilon^{21}Q_{21}$. The other two will not leave the solution invariant but generate supertranslational modes. Firstly we consider the scalar equations:
\begin{equation}
r \partial_r \phi_1 = + f \phi_1,\quad r \partial_r \phi_2 + \phi_2 = + f \phi_2
\label{BPS1}
\end{equation}

They are very similar in nature, differing only in the linear part, which means they can potentially be related to each other by the right transformation. It is in fact the case: if $\phi_1$ obeys its equation of motion then we are free to take
\begin{equation}
	\phi_2 = \frac{\rho}{r}\phi_1\equiv  \frac{\rho}{r}\phi
\end{equation}
for some unknown constant length scale $\rho$, to obtain a solution to the second BPS scalar equation. This new length scale, the size modulus, defines a new regime for the spreading of the solutions, and is responsible for the semi-local nature of the vortex. As a consequence of its appearance, the various fields constituting the vortex will decay as rational functions of $r,\rho$. The single undetermined scalar profile function inside $q^{1}$ is then relabeled $\phi$. We will see later on in Eq.(\ref{explicitsol}) that this parameter is exactly analogous to the $\mathbb{CP}(1)$ instanton lump size modulus that our solution must at some level reproduce, given the vacuum manifold.

The sign of the right hand side is fixed by the supercharges we fixed as well as the requirement that the profiles introduced in the Ansatz are positive. Then,  $\phi$ should be regular at the origin and reach the vacuum expectation value at infinity, $\phi$ is an increasing function of $r$, $f$ being positive in our Ansatz confirms this.

In addition the BPS equations also produce the following constraint for the gauge profile $f$:
\begin{equation}
-\frac{1}{r}\partial_r f +g^2\left(\phi^2\left(1+\frac{\rho \bar{\rho}}{r^2}\right) -\xi \right)  =0\,.
\label{BPS4}
\end{equation}

Immediately, this allows us to write super-translational  zero modes for the theory. They are generated by $\epsilon^{11}Q_{11},\,\epsilon^{22}Q_{22}$, which act non-trivially on the BPS string solution, enabling us to use the BPS equations to simplify the zero modes:
\begin{align}
\delta \bar{\psi}^{1}_{\dot{2}}&= i\sqrt{2}\bar{\slashed{D}}_{\dot{2}1}\bar{q}_{A}\epsilon^{11}=-2\sqrt{2}\left( \frac{x+iy}{r^2}\right) f(r)
\phi(r) \epsilon^{11}\,, \nonumber\\
\delta \bar{\psi}^{2}_{\dot{2}}&= i\sqrt{2}\bar{\slashed{D}}_{\dot{2}1}\bar{q}_{A}\epsilon^{11}=+2\sqrt{2}\left( \frac{x+iy}{r^2}\right) \left(1-f(r)\right) 
\phi(r) \frac{\bar{\rho}e^{i\theta}}{r}
\epsilon^{11} \,,\nonumber\\
\delta \bar{\tilde{\psi}}^{1}_{\dot{1}}&=i\sqrt{2}\bar{\slashed{D}}_{\dot{1}2}\bar{q}_{A}\epsilon^{22}=2\sqrt{2}\left( \frac{x-iy}{r^2}\right) 
\phi(r)f(r) 
 \epsilon^{22} \,,\nonumber \\
\delta \bar{\tilde{\psi}}^{2}_{\dot{1}}&=i\sqrt{2}\bar{\slashed{D}}_{\dot{1}2}\bar{q}_{A}\epsilon^{22}=-2\sqrt{2}\left( \frac{x-iy}{r^2}\right) 
\left(1-f(r) \right) \phi(r) \frac{\rho e^{-i\theta}}{r}
 \epsilon^{22} \,,\nonumber \\
\delta\lambda^{11} &=+2D^3 (\tau^{3})^{1}_{1} \epsilon^{11}= -2i g^2\left( \phi^2 \left( 1 + \left|\frac{\rho}{r}\right|^2 \right) - \xi\right) \epsilon^{11}\,,\nonumber\\
\delta\lambda^{22} &= -2D^3 \left( \tau^{3}\right) ^{1}_{1} \epsilon^{22}= +2i g^2 \left( \phi^2\left( 1 + \left|\frac{\rho}{r}\right|^2\right)   - \xi\right) \epsilon^{22}\,.
\label{stzm}
\end{align}
All others are identically zero, by satisfaction of the BPS equations. The fermions $\epsilon^{11},\,\epsilon^{22}$ can be turned into dynamical worldsheet variables, we preemptively label them respectively $\zeta_L,\,\zeta_R$. They are the fermionic superpartners on the worldsheet of the translational zero mode of the vortices.  

The second set of zero modes are generated by $\epsilon^{12},\,\epsilon^{21}$, which usually act trivially on the string solution. However, adding slow variations of $\rho$ in $(t,z)$ changes this: then, we can write zero modes depending on derivatives of $\rho$. Computing them requires a bit more effort, since in this case the fermionic parameters connect in an unobvious way to the associated worldsheet dynamical fermions, as opposed to the previous case. For starters, we need to updated our gauge Ansatz: in order to retain gauge invariance, new components of the gauge field are required to be turned on. For $k=(t,z)$
\begin{equation}
 A_k=-i(\rho^*\partial_k \rho - \rho \partial_k \rho^*)\gamma(r)	
\end{equation}
which introduces a new radial profile function $\gamma(r)$, constrained by the gauge equations of motion.

In the case of non-Abelian vortices, where a similar analysis was conducted leading to super-orientational modes, this extra gauge profile function was solved for explicitly by studying its equation of motion, and an exact solution was found in terms of the profile $\phi$ alone. This is not as easy in the present case, since the $\rho$ modulus intervenes in every radial profile in the Ansatz, the minimisation equation is much more complicated. In a previous work, a complete solution was found in the low energy limit by sending $m_W=g^2\sqrt{\xi}$ to infinity, or, more precisely, by placing oneself sufficiently far from the core whose width is defined by $(g^2\sqrt{\xi})^{-1}$. Then, the solution takes the following form \cite{Achucarro:1999it,Shifman:2011xc}:
\begin{equation}
	\phi(r)=\frac{\sqrt{\xi}r}{\sqrt{r^2+\bar{\rho}\rho}},\quad f(r)=\frac{|\rho|^2}{r^2+|\rho|^2},\quad\gamma(r)=\frac{1}{2\left(r^2+|\rho|^2 \right) }
	\label{explicitsol}
\end{equation}
The structure is indeed as predicted very similar to the $\mathbb{CP}(1)$ instanton, far away. The semi-local nature of the vortices is clearly seen at this distance from the core, but it is only an approximate solution to the various equations at hand. In particular it leaves the gauge BPS equation (\ref{BPS4}) somewhat vacuous: while the matter equation is solved exactly by this solution, the gauge one is only approximately solved, it is an asymptotic solution. In order to ensure we write precise statements and algebraic relations, we would like to stick to exact, if implicit, profile solutions. 

\subsection{Modulus Fluctuations: Holomorphy Equations}
Surprisingly it is possible to find a great deal of information about the implicit solutions for $\gamma$, so long as we impose holomorphy of SUSY variations. Assuming nothing about the function $\gamma$, we can write out the full SUSY variations of the matter and gauge fermions under transformations with parameters $\epsilon^{12},\,\epsilon^{21}$ when we assume that $\rho$  is no longer a constant modulus, but actually has a dependence on $(t,z)$. The resulting variations are of course no longer vanishing, in general they are a function of both $\partial \rho$ and $\partial \bar{\rho}$. 

Constraints on $\gamma$ then occur when we impose that these transformations should be \textit{holomorphic}: after a fermionic variation we expect the fermionic zero mode to only depend on exactly one of  $\partial \rho$ and $\partial \bar{\rho}$, not both. One may note that this simplification already happens when using the approximate but explicit solutions detailed in Eq.(\ref{explicitsol}).

  The simplest case is the variation of $\bar{\tilde{\psi}}^{2}$, since it involves the field $q^2$: it already has a very direct dependence on $\rho$ and not its conjugate, so we expect its variation should be proportional to $\partial \rho$ only. This gives us a first constraint on $\gamma$: with the sign of $A_{t,z}$ chosen above, we have
\begin{equation}
	\partial_{|\rho|^2}\phi=-\gamma\phi
	\label{BPS2}
\end{equation}
We expect $\phi$ to decrease with $\rho$ since it is a size modulus, it controls the spreading of the profile in space. This is consistent with choosing $\phi,\,\gamma$ positive. This assumption then produces a holomorphic dependence on $\partial \rho$ for $\bar{\tilde{\psi}}^{1}$, which is obvious since those two fields were already related by a previously-used BPS equation.

Secondly, let us also observe what additional conditions are imposed from the gaugino supersize zero mode, we obtain a second equation for $\gamma$:

\begin{equation}
	\partial_{|\rho|^2} f = \pm r \partial_r \gamma
	\label{BPS3}
\end{equation}
where the sign controls which of the two zero modes depend on $\partial \rho$, the other zero mode will depend on the conjugate. Unlike in the matter case, there is no good heuristic to determine which needs to be true. As it happens, however, while we \textit{a priori} could pick either, this choice is actually forced onto us: indeed, the scalar BPS equation (\ref{BPS1}) and the scalar holomorphy equation (\ref{BPS2}) generate this third one, as can be seen by expressing $\partial_r \partial_{|\rho|^2} \phi$ in two different but equal ways. 

The choice of sign in previous cases dictates that this sign should be negative. Furthermore, physically, $f$ is also expected to increase with $\rho$, since this gauge field should vanish for vanishingly small $\rho$, whereas $\gamma$ is expected to decrease with $r$. Again it can be noted that the explicit solutions for the profiles satisfy these holomorphy relations exactly:
\begin{equation}
\partial_{|\rho|^2} f = - r \partial_r \gamma
\end{equation}

By differentiating the gaugino BPS equation (\ref{BPS4}) by $|\rho|$, and using the newly generated identities involving $\gamma$, we obtain precisely the equation of motion for this additional profile, obtained by substituting $A_k$ in the action directly \cite{Shifman:2011xc}:
\begin{equation}
		\frac{1}{r}\partial_r\left( r \partial_r \gamma\right) + g^2\left(-2\phi^2\left(1+\frac{\rho \bar{\rho}}{r^2}\right) \gamma + \frac{\phi^2}{r^2} \right)  =0
		\label{rhomin}
\end{equation}

Thus, we have proven that using the above first-order equations, both the BPS and the holomorphy equations, we are in principle obtaining a solution to the $\gamma$ equations of motion. Again one notices that the explicit solution (\ref{explicitsol}) satisfies this equation only asymptotically.

\subsection{Computing Supersize Zero Modes}
Once this is done,  the supersize zero modes are of the correct form for interpretation as being proportional to worldsheet fermion zero modes. This is, every fermionic parameter comes multiplied with one of $\partial \rho$ or $\partial \bar{\rho}$, which, using the worldsheet SUSY variations, can then be wholly replaced by a worldsheet fermion zero mode, see review \cite{SYrev} where similar procedure was used for calculating superorientational fermionic zero modes for  non-Abelian string.

In total, we get the following expressions for the super-size modes:
\begin{align}
\delta \bar{\tilde{\psi}}^{2}_{\dot{1}}&=+i\sqrt{2}\epsilon^{12}\frac{e^{-i\theta}}{r}\left(2\bar{\rho}\rho \gamma(r) -1 \right)\phi(r) \left(\partial_0 + i \partial_3 \right) \rho  \,,\nonumber\\
\delta \bar{\psi}^{2}_{\dot{2}}&=- i\sqrt{2}\epsilon^{21}\frac{e^{+i\theta}}{r}\left(2\bar{\rho}\rho \gamma(r) -1 \right)\phi(r) \left(\partial_0 - i \partial_3 \right) \bar{\rho} \,,\nonumber\\
\delta \bar{\tilde{\psi}}^{1}_{\dot{1}}&= i\sqrt{2}\epsilon^{12}\left(2\gamma(r) \phi(r) \bar{\rho} \right) \left(\partial_0 + i \partial_3 \right) \rho \,,\nonumber \\
\delta \bar{\psi}^{1}_{\dot{2}}&=-i\sqrt{2}\epsilon^{21}\left(2\gamma(r) \phi(r) \rho \right) \left(\partial_0 - i \partial_3 \right) \bar{\rho}\,,\nonumber\\
\delta \lambda^{11} &=  +2\frac{\epsilon^{21}(x-iy)\gamma(r)}{r}\rho\left(\partial_0 - i \partial_3 \right)\bar{\rho} \,,\nonumber\\
\delta \lambda^{22} &= -2\frac{\epsilon^{12}(x+iy)\gamma(r)}{r}\bar{\rho}\left(\partial_0 + i \partial_3 \right)\rho\,.
\label{sszm1}
\end{align}

These solutions can then be further simplified by substituting for the derivatives of $\rho$ using worldsheet supersymmetry: construct fermionic zero modes on the worldsheet from the variations of $\rho$. We introduce two spinor-valued parameters $\eta,\, \xi$ to generate two SUSY transformations:
\begin{equation}
\delta \chi_\alpha=i\sqrt{2}\slashed{\partial}_{\alpha\beta}\rho \left( \eta^\beta + i \xi^\beta\right),\quad \delta \bar{\chi}_\alpha=i\sqrt{2}\slashed{\partial}_{\alpha\beta}\bar{\rho} \left( \eta^\beta -i \xi^\beta\right)
\end{equation}
or in components
\begin{align}
\delta \chi_R&=i\sqrt{2}\left(\partial_0 + i \partial_3 \right) \rho \left( \eta^2 + i \xi^2\right),\quad \delta \bar{\chi}_R=i\sqrt{2}\left( \partial_0 + i \partial_3\right) \bar{\rho} \left( \eta^2 -i \xi^2\right)\,,\nonumber\\
\delta \chi_L&=i\sqrt{2}\left(\partial_0 - i \partial_3 \right) \rho \left( \eta^1 + i \xi^1\right),\quad \delta \bar{\chi}_L=i\sqrt{2}\left( \partial_0 - i \partial_3\right) \bar{\rho} \left( \eta^1 -i \xi^1\right)\,.
\end{align}
This sign convention reflect the fact we are in Euclidean space.

Thus, we identify $\epsilon^{12}=(\eta^2+i\xi^2)$ and $\epsilon^{21}=(\eta^1 - i\xi^1)$, enabling us to write the final form of the supersize zero modes:

\begin{align}
\delta \bar{\tilde{\psi}}^{2}_{\dot{1}}&=+\frac{e^{-i\theta}}{r}\left( \left(2\bar{\rho}\rho \gamma(r) -1 \right)\phi(r)\right)  \delta\chi_R \,,\nonumber\\
\delta \bar{\psi}^{2}_{\dot{2}}&=- \frac{e^{+i\theta}}{r}\left( \left(2\bar{\rho}\rho \gamma(r) -1 \right)\phi(r)\right)  \delta \bar{\chi}_L\,,\nonumber\\
\delta \bar{\tilde{\psi}}^{1}_{\dot{1}}&= +\left(2\gamma(r) \phi(r) \bar{\rho} \right)  \delta\chi_R\,,\nonumber \\
\delta \bar{\psi}^{1}_{\dot{2}}&=-\left(2\gamma(r) \phi(r) \rho \right)  \delta \bar{\chi}_L\,,\nonumber\\
\delta \lambda^{11} &=  -i\sqrt{2}\frac{(x-iy)\partial_r\gamma(r)}{r}\rho \delta \bar{\chi}_L \,,\nonumber\\
\delta \lambda^{22} &= +i\sqrt{2}\frac{(x+iy)\partial_r\gamma(r)}{r}\bar{\rho} \delta\chi_R\,.
\label{sszm2}
\end{align}

Inserting these into the spacetime action, one readily gets kinetic terms for these fermions, forming a full $(2,2)$ sigma model on the worldsheet. In order to define useful normalisation constants due to integration over the transverse spacetime, let us quickly check the form of this Lagrangian.

\subsection{$(2,2)$ Supersymmetric Worldsheet Elements}

First we compute the kinetic term for the size modulus, which involves integrating over the profiles. We again come to some simplifications when using the first order equations (\ref{BPS2}),(\ref{BPS3}),(\ref{BPS4}) and the minimisation equation (\ref{rhomin}). Indeed, we get two terms that contribute to a kinetic term for $\rho$: one from the gauge field and one from the scalars. From the former we have
\begin{equation}
	\frac{1}{g^2}F_{ik}F_{ik}=\frac{4\bar{\rho}\rho}{g^2}\left( \partial_r\gamma\right)^2=-2\bar{\rho}\rho\gamma\frac{2}{g^2}\left(\frac{1}{r}\partial_r \left( r\partial_r\gamma \right) \right)^2 +(\text{total derivative}),
\end{equation}
 and from the latter 
\begin{align}
	\left(Dq_i \right)^\dagger \left(Dq_i \right)&=\frac{\phi^2}{r^2} + 4\bar{\rho}\rho \gamma\frac{\phi^2}{r^2}\left(-1+r^2\gamma + \bar{\rho}\rho\gamma \right) \nonumber\\
	&=\frac{\phi^2}{r^2} + 2\bar{\rho}\rho \gamma\left(2\phi^2\gamma\left( 1+\frac{\bar{\rho}\rho}{r^2}\right)   - 2\frac{\phi^2}{r^2} \right)\,.
\end{align}
We have written both these components conspicuously in order to make apparent the terms that also appear in Eq.(\ref{rhomin}). Summing these two and applying the minimisation condition, the full integral which produces the $\rho$ kinetic term simplifies massively and we obtain
\begin{equation}
	\mathcal{L}_{\rho,\text{kin.}}=I\left( \partial \rho \partial \bar{\rho}\right) =\left( 2\pi\int r\,dr\left( \frac{\phi^2}{r^2}\left( 1 -2\bar{\rho}\rho\gamma\right) \right) \right) \left( \partial \rho \partial \bar{\rho}\right) 
\end{equation}
We can check that this produces the right result by inserting the explicit solution \ref{explicitsol}. The integrals this produces is divergent, the field profiles do not decay fast enough at large $r$. We impose an infrared cutoff, integrating only up to a large length $L_{IR}$ in the plane transverse to the string, the integral produces, cf. \cite{Shifman:2006kd}
\begin{align}
	I&=2\pi\int_0^L dr\left(\frac{\xi r}{r^2 + \bar{\rho}\rho} \right) \left(1-\frac{\bar{\rho}\rho}{r^2+\bar{\rho}\rho} \right)\nonumber\\ &=\pi\xi\left(\log\left(1+\frac{L_{IR}^2}{\bar{\rho}\rho} \right) - \frac{L_{IR}^2}{L_{IR}^2+\bar{\rho}\rho}  \right) \sim \pi\xi \log\left( \frac{L_{IR}^2}{\bar{\rho}\rho}\right),
\end{align}
where we consider infra-red (IR) logarithm $\log{(L_{IR}/|\rho|)}\gg 1$ as a large parameter. Clearly, the IR
logarithm here comes from the profile function of the  second massless flavor, see Sec. 2.

Note, that  modes with IR logarithmically
divergent norms are on the borderline between normalizable and non-normalizable modes. Usually such modes are
considered as “localized” on the string, while   power non-normalizable modes are associated with vacuum rather then with a string.  We follow this rule and include modulus $\rho$ in our effective world sheet theory on the string, see \cite{Shifman:2006kd,Shifman:2011xc}.

The metric on $\rho$ is K\"{a}hlerian, originating from the following potential
\begin{equation}
	\mathcal{K}(\rho,\bar{\rho})=\bar{\rho}\rho \log\left( \frac{L_{IR}^2}{\bar{\rho}\rho}\right) 
	\label{kpot}
\end{equation}
Note, that with logarithmic accuracy we do not differentiate the IR logarithm.

An entirely analogous computation with the fermionic supersize modes produces the exact same kinetic normalisation for the worldsheet fermions, agreeably.
\begin{equation}
\pi\xi \log\left( \frac{L_{IR}^2}{\bar{\rho}\rho}\right)\,   i \left(\bar{\chi}_R \partial_L \chi_R +  \bar{\chi}_L \partial_R \chi_L \right).
\end{equation}
Note that IR logarithm here comes from $1/r$ tails of massless fermions $\psi^2$, $\tilde{\psi}^2$ of the second flavor in \eqref{sszm2}, while massive fermions of the first flavor have faster decay at infinity and do not produce IR logarithms, see Sec 2. 

Thus, the \ntwot supersymmetric world sheet theory on the string at $\mu=0$ reads 
\beqn
S_{2D} &=& \int d^2 x \;\pi\xi \left\{\log\left( \frac{L_{IR}^2}{|\rho|^2}\right)\,\left[ |\pt_k \rho|^2 +
 i \bar{\chi}_R \partial_L \chi_R + i \bar{\chi}_L \partial_R \chi_L \right] 
\right.
\nonumber\\[3mm]
&+&
\left.
|\pt_k x^i|^2
+ i \bar{\zeta}_R \partial_L \zeta_R + i \bar{\zeta}_L \partial_R \zeta_L  \right\}
\label{wcN=2}
\eeqn
with logarithmic accuracy, $k=0,3$ label the world sheet coordinates. Here we included also translational modes 
$x^i$, $i=1,2$ and their superpartners 
$\zeta_{L}$, $\zeta_{R}$, see \eqref{stzm}. We see that translational and size sectors do not interact.
We will see later that this will change once we switch on $\mu$ deformation.

\section{Deforming the Spacetime Theory}
\label{sec2}
\setcounter{equation}{0}

Now that the worldsheet theory has been created, we are able to observe how it responds to modifications of the spacetime theory. Specifically, we have enough supercharges to allow a further partial breaking of supersymmetry, while still retaining a supersymmetric worldsheet as an end product. Let us see how this happens.

We now add a SUSY breaking superpotential \eqref{msuperpotbr} to the spacetime theory to produce an $\mathcal{N}=1$ Lagrangian.
It gives a mass term to the gauge scalar $a$ and one of the gauginos $\lambda^{\alpha 2}$, which form a SUSY doublet $\mathcal{A}$. Upon taking the large $\mu$ limit, this decouples the extra adjoint fields and one gets a theory similar to $\mathcal{N}=1$ SQED, with extra flavour and particular charges. This potential preserves $\epsilon^{11}Q_{11},\,\epsilon^{21}Q_{21}$ so that the string solution now only has two supercharges left, generated by $\epsilon^{21}Q_{21}$ and its conjugate. 

However, the other charges still generate fermionic zero modes, for small $\mu$ at least. By general considerations on index theorems a small deformation of this kind cannot cause fermion zero modes to drop out of the spectrum. 

Though still existent, the fermionic zero modes are affected by these modifications. Those proportional to the parameters preserved by the addition of this $\mu$ term do not change. Thus, both in the supertranslational case in Eq.(\ref{stzm}) and the super-size case in Eq.(\ref{sszm2}), $\delta\bar{\psi}_{\dot{2}}$ and $\delta\lambda^{11}$ (proportional to $\epsilon^{11}=\zeta_L$ or $\epsilon^{21}\propto \bar{\chi}_L$) do not change, while
$\delta\bar{\tilde{\psi}}_{\dot{1}}$ and $\delta\lambda^{22}$ (proportional to $\epsilon^{22}=\zeta_R$ or $\epsilon^{12}\propto \bar{\chi}_R$) get modified profiles that become $\mu$ dependent. By analysing the Dirac equation, it is possible to find approximate solutions for these profiles respectively as a perturbation series in $\mu$ for small values thereof.

The modifications of these profiles make the fermion zero modes overlapping, thus causing interactions between supertranslational and supersize modes, and creating a general $\mathcal{N}=(0,2)$ worldsheet theory that does not benefit from any supersymmetry enhancement. This kind of enhancement is especially easy to fall into in our case. Indeed, any supersymmetric NLSM whose target space is a K\"{a}hler manifold is automatically $\mathcal{N}=(2,2)$, which we referred to as Zumino's theorem. Since the target spaces for both of our basic coordinates, the translational mode $(y\pm iz)$ and the size mode $\rho,\,\bar{\rho}$, are both complex one-dimensional manifolds, they are automatically K\"{a}hler (the K\"{a}hler form is necessarily closed as it is a top-form). The most sure-fire way to ensure no enhancement occurs accidentally is then to couple fermionic variables from both target spaces together, without, of course, changing the structure of the bosonic coordinates i.e. deforming the manifold itself.

\subsection{Dirac Equations for Spacetime Fermions}

Once the theory is deformed by the potential we added, fermionic zero modes in the theory will generically not be holomorphic anymore. That is, they may depend on a worldsheet spinor and on its conjugate, and in different ways at that. In this spirit we suggest writing the fermionic zero modes in a generic form, with arbitrary profile functions, for which the Dirac equation then provides a constraint. The full Dirac equations are
\begin{align}
	&\frac{i}{g^2} \left( \slashed{{D}}\bar{\lambda}\right)^{f}_{\dot{\alpha}} +{i\sqrt{2}} \left( {\psi}^A_{\dot{\alpha}} \bar{q}^{Af} + q^{Af} {\tilde{\psi}}^A_{\dot{\alpha}} \right) -\mu \delta^f_2 {\lambda}^2_{\dot{\alpha}}=0\,,\nonumber\\
	&i \left( \slashed{D}\bar{\psi}\right)^\alpha +i\sqrt{2}\bar{q}_f\lambda^{\alpha f}  =0,\quad i \left( \slashed{D}\bar{\tilde{\psi}}\right)_\alpha +i\sqrt{2}q^f\lambda_{\alpha f}  =0\,.
\end{align}

%
%

Convenient parametrisations for the modified profiles are the following. For the supertranslational modes:

\begin{align}
	\lambda^{22}&=\lambda_0(r) \zeta_R + \lambda_1(r) \frac{x+iy}{r}\bar{\zeta}_R \,,\nonumber\\
	\bar{\tilde{\psi}}^1_{\dot{1}}&=\left(\frac{x-iy}{r}\psi^1_0(r)\zeta_R +\psi^1_1(r)\bar{\zeta}_R \right)\,,\nonumber\\
	\bar{\tilde{\psi}}^2_{\dot{1}}&=\frac{\rho e^{-i\theta}}{r}\left(\frac{x-iy}{r}\psi^2_0(r)\zeta_R +\psi^2_1(r)\bar{\zeta}_R \right)\,.
\end{align}
This produces the following profile equations

\begin{align}
	&\partial_r \lambda_0 -ig^2\sqrt{2} \phi \left(\psi^1_0+\frac{\bar{\rho}\rho}{r^2} \psi^2_0\right) -g^2 \mu \lambda_1=0 \,,\nonumber \\
	&\partial_r \lambda_1 +\frac{\lambda_1}{r} - ig^2\sqrt{2} \phi \left(\psi^1_1+\frac{\bar{\rho}\rho}{r^2}\psi^2_1 \right)\psi^1_1-g^2 \mu \lambda_0=0\,,\nonumber\\
	&\partial_r \psi^{1,2}_0 +\frac{1}{r}\psi^{1,2}_0\left( 1-f\right) -i\sqrt{2}\phi\lambda_0=0\,,\nonumber\\
	&\partial_r \psi^{1,2}_1 -\frac{f}{r}\psi^{1,2}_1 -i\sqrt{2}\phi\lambda_1=0\,.
	\label{trmodeeqs}
\end{align}

For the super-size modes we propose the following parametrisation:

\begin{align}
	\lambda^{22}&=\frac{x+i y}{r}\lambda_+(r)\bar{\rho}\chi_R   + \lambda_-(r)\rho \bar{\chi}_R \,,\nonumber\\
\bar{\tilde{\psi}}^1_{\dot{1}}&=\left(\psi^1_+(r)\bar{\rho}\chi_R +\frac{x-iy}{r}\psi^1_-(r) \rho\bar{\chi}_R \right)\,,\nonumber\\
\bar{\tilde{\psi}}^2_{\dot{1}}&=\frac{ e^{-i\theta}}{r}\bar{\rho} \rho\left(\psi^2_+(r)\chi_R  +\frac{x-iy}{r}\psi^2_-(r) \frac{\rho}{\bar{\rho}}\bar{\chi}_R\right)\,.
\end{align}
leading to the following profile constraints:
\begin{align}	
&\partial_r \lambda_+ + \frac{\lambda_+}{r} -ig^2\sqrt{2}\left(\psi^1_+ +\frac{\bar{\rho}\rho}{r^2}\psi^2_+ \right)\phi  -g^2\mu\lambda_- = 0 \,,\nonumber \\	
&\partial_r \lambda_- -ig^2\sqrt{2}\left(\psi^1_- +\frac{\bar{\rho}\rho}{r^2}\psi^2_- \right)\phi  -g^2\mu\lambda_+ = 0 \,,\nonumber \\	
&\partial_r \psi^{1,2}_+ -\frac{f}{r}\psi^{1,2}_+ -i\sqrt{2}\phi\lambda_+=0\,,\nonumber\\
&\partial_r \psi^{1,2}_- +\frac{1}{r}\psi^{1,2}_-\left( 1-f\right) -i\sqrt{2}\phi\lambda_-=0\,.
\label{sizemodeeq}
\end{align}

These parametrisations were chosen to satisfy several conditions: one, they should capture features present when $\mu=0$ (particularly complex phases and singularities), two, the matter profiles should be scalars of consistent mass dimension, three, the profiles should be invariant under phase rotations affecting $\rho$ and its superpartner.

\subsection{Small $\mu$ Solutions}

The equations obtained at small $\mu$ can be solved order by order. The $(+)$ and $(0)$ profiles are the only ones that survive taking $\mu\rightarrow 0$, so these profiles will only have even powers of $\mu$ whereas the $(-)$ and $(1)$ profiles will capture all the odd powers of $\mu$. The Dirac equation then couples these two together in a consistent, order by order expansion. 

Thus, we can start off by writing the $(+)$ and $(0)$ profiles at zeroth order, from which we can compute the others. This gives us, in the translational case:
\begin{align}
	\lambda_0 = 2ig^2 \left( \phi^2\left(1+\frac{\bar{\rho}\rho}{r^2} \right)-\xi \right) ,\quad \psi^1_0=2\sqrt{2}\frac{f \phi}{r},\quad \psi^2_0=2\sqrt{2}\frac{(f-1) \phi}{r}\,.
\end{align}
By virtue of the BPS equations, these profiles are a solution to the Dirac equations above for vanishing $\mu$. We then use these to source the equations for $\lambda_1,\, \psi_1$: given the high degree of similarity between the $(0)$ and $(1)$ equations, differing only by terms linear in the profile functions, we try a solution of the form \begin{equation}
\lambda_1=b(r)\lambda_0,\quad \psi_{1,2}=b(r)\psi_0
\end{equation} for some unknown function $b$. The equations for the $(1)$ profiles subsume to two condition on $b$, notably
\begin{equation}
	\partial_r b +\frac{b}{r} +\mu g^2 =0,\quad  \partial_r b -\frac{b}{r}=0\,.
\end{equation}
This is solved by $b(r)=-\frac{\mu g^2r}{2}$. The $(1)$ profiles are therefore

\begin{align}
	 &\lambda_1 =-i\mu g^4r \left( \phi^2\left(1+\frac{\bar{\rho}\rho}{r^2} \right)-
	 \xi \right),\nonumber\\
	  &\psi^1_1=-\sqrt{2}\mu g^2f \phi,\quad \psi^2_1=-\sqrt{2}\mu g^2(f-1) \phi\,.
\end{align}
This is entirely analogous to the local non-Abelian case.

For the supersize moduli, the $(+)$ profiles at zeroth order are
\begin{align}
	\lambda_+ =i\sqrt{2}\partial_r \gamma,\quad  \psi^1_+=2\gamma \phi ,\quad  \psi^2_+=2\gamma \phi - \frac{\phi}{\bar{\rho}\rho}\,.
\end{align}
The zeroth order equation for the $(+)$ profiles subsumes to the extremisation equation for $\gamma$. Thanks to our parametrisation, we can apply the same kind of trick again to find the $(-)$ profiles: writing 
\begin{equation}
\lambda_-=-\frac{\mu g^2r}{2}(\lambda_+-i \sqrt{2} c(r)),\,\psi^{1,2}_-= -\frac{\mu g^2r}{2} \psi^{1,2}_+,
\end{equation}
we obtain a solution to the Dirac equation when 
\begin{equation}
	c=-\frac{2}{r}\gamma\,.
\end{equation}
This gives the following profiles:

\begin{align}
	&\lambda_-=-\frac{i\mu g^2 \sqrt{2}r}{2}\left(\partial_r \gamma + \frac{2}{r}\gamma \right),\nonumber\\
	 &\psi^{1}_- = -\mu g^2r \gamma \phi ,\quad  \psi^2_- = -\mu g^2r \left( \gamma \phi-\frac{\phi}{2 \rho \bar{\rho}}\right) \,.
\end{align}

With these profiles supersize zero modes can be checked to  be non-singular at zero 
and normalizable at infinity to the order $O(\mu)$ by using the explicit solution \eqref{explicitsol},
up to a caveat we detail in \ref{ap2}.

We can now feed these profiles into the kinetic terms of the 4d fermions and observe any mixing between worldsheet modes. At this level we can expect three changes to occur, three coefficients that can depart from their expected value. The changes affect the $\zeta_r,\,\chi_R$ worldsheet fermions, so their respective kinetic terms can change normalisation: label them $I_{\zeta\zeta},\,I_{\chi\chi}$. But also, we expect a mixing term between these two fields to occur: if the shape of the interactions persists to be K\"{a}hlerian then Zumino's theorem will apply and one would observe an enhancement of the number of supersymmetries.

If $\mu=0$ this coefficient vanishes, since, for instance, $\lambda_0$ and $\lambda_+$ have no overlap. One comes multiplied by $\frac{x+i y}{r}$ while the other does not, similarly for the matter fermions.

It is clear that at leading order in $\mu$, the fermion kinetic constant for $\zeta$ does not change from its initial value, which one can show is the integral of a total derivative by using the Maxwell equation

\begin{equation}
	I_{\zeta \zeta} = \int rdrd\theta \quad \left( \left( \frac{1}{r}\partial_r f(r)\right)^2 +\frac{1}{r} J\right)   = \left[ \frac{1}{r}f(r)\partial_r f(r) \right]^\infty_0 =1 \,.
\end{equation}
For precisely the same reasons, at order $O(\mu)$ the $\chi$ normalisation does not change either. In that case, a caveat must be raised, the details of which are in Appendix \ref{ap2}.

Now, with these solutions, it is the case that zero modes from the translational and size moduli are able to mix, leading to the sought-after term on the worldsheet:
\begin{equation}
 { \pi g^2\xi\mu} \log\left( \frac{L_{\text{IR}}^2}{\bar{\rho}\rho}  \right) \left(\zeta_R \chi_R \partial_L\bar{\rho} +\text{c.c.} \right)  \label{suggest}
\end{equation}
where  we keep only terms which contain IR logarithms. Here again the IR logarithm comes from the massless
fermion of the second flavor.

The shape of this resulting term is in fact fixed by supersymmetry and target space invariance, as we will see in the next section. In obtaining this expression, we again used the fact that radial variations of $\rho$ are negligible, since they occur systematically in comparison to $L_{IR}$. This enabled  us to justify treating the logarithmic factors in the kinetic terms as constants and changing normalisation to remove them, here it enables us to write
\begin{equation}
\bar{\rho}\partial \rho \approx -\rho \partial {\bar{\rho}}
\end{equation}
which simplifies the computation to the result quoted above.

Thus our world sheet theory to the $O(\mu)$ order becomes
\begin{align}
S_{2D} &= \int d^2 x \;\pi\xi \left\{\log\left( \frac{L_{\text{IR}}^2}{|\rho|^2}\right)\,\left[ |\pt_k \rho|^2 +
 i \bar{\chi}_R \partial_L \chi_R + i \bar{\chi}_L \partial_R \chi_L \right] 
\right.
\nonumber\\
&+
\left.
 i \bar{\zeta}_R \partial_L \zeta_R + g^2\mu \log\left( \frac{L_{\text{IR}}^2}{|\rho|^2}  \right) \left(\zeta_R \partial_L \bar{\rho}\chi_R  +\text{c.c.} \right) \right\},
\label{wcN=1}
\end{align}
where we drop translational moduli $x^i$ and $\zeta_L$ which are sterile.

The mixing term, by its existence, breaks $\mathcal{N}=2$ supersymmetry, as has been discussed. 
Absorbing with logarithmic accuracy square roots of IR logarithms in the normalization  for $\chi_R$ and $\rho$  we finally arrive at the action
\begin{align}
S_{2D} &= \int d^2 x \;\pi\xi \left\{ |\pt_k \rho|^2 +
i \bar{\chi}_R \partial_L \chi_R + i \bar{\chi}_L \partial_R \chi_L 
\right.
\nonumber\\
&+
\left.
i \bar{\zeta}_R \partial_L \zeta_R + g^2\mu  \left(\zeta_R  \partial_L\bar{\rho} \chi_R +\text{c.c.} \right) \right\}.
\label{wcN=1final}
\end{align}

We see that the mixing term  also does not contains IR logarithm and  becomes of order $g^2\mu$. 

As we mentioned, the shape of this term is expected from supersymmetry: there exists a specific way of combining $(0,2)$ superfields in such a way as to generate a mixing term of this form, but the above result is not the complete answer:  along with this new term, extra four-fermion interactions are generated due to $F$ terms. In order to determine the full expression, let us turn to this formalism to generate the remainder of the Lagrangian.

\subsection{Superspace Action}
We have found that the worldsheet theory develops a deformation term that breaks $(2,2)$ supersymmetry. This term mixes fermions living in different target spaces, while the bosonic coordinates of the manifolds do not mix. Evidence of leftover supersymmetry after this breaking is most easily seen by writing a $(0,2)$ superfield formulation of the Lagrangian.

 We introduce three superfields, whose expansions in chiral superspace coordinates are
\begin{equation}
	A=\rho + \theta \sqrt{2}\chi_L ,\quad B=\chi_R + \sqrt{2}\theta F_s,\quad C=\zeta_R +\sqrt{2}\theta F_t
\end{equation}
where $F_t,\,F_x$ are unimportant auxiliaries leading to four-fermion interactions. We also introduce the K\"{a}hler 1-forms $\mathcal{K}_{z},\mathcal{K}_{\bar{z}}$, which are complex conjugate and arbitrary. They would derive, in a $(2,2)$ setting expressed in $(0,2)$ notation, from the K\"{a}hler potential by 
\begin{equation}
	\mathcal{K}_{z} = \partial_z \mathcal{K}\,.
\end{equation}
We then define the metric of the space by $G_{z\bar{z}}=\partial_{\bar{z}} \mathcal{K}_{z}=\overline{{G_{\bar{z}z}}}$.

 The $\mathcal{N}=(2,2)$ Lagrangian, written in this $(0,2)$ language, takes the following form, first in a generic formulation and then in our specific case:
 
 \begin{align}
 \mathcal{L}_{(2,2)}=&\pi\xi\int d^2\theta\, \left(i	\mathcal{K}_{z} 
\partial_R A + \text{c.c.}
+ G_{z\bar{z}} B^\dagger B  \right) \\
=&\pi\xi\int d^2\theta\, \left(i \log\left( \frac{L_{IR}^2}{ A^\dagger A}\right) \left(A^\dagger\partial_R A - \partial_R A^\dagger A \right)\right. \nonumber\\
&\left.+  \log\left( \frac{L_{IR}^2}{ A^\dagger A}\right) B^\dagger B  \right) 
\end{align}
given that the K\"{a}hler potential was given in Eq.(\ref{kpot}). Then, a term that explicitly breaks $(2,2)$ supersymmetry can be found by coupling $B$ and $A^\dagger$ directly, without involving $A$. The following term is suitable:
\begin{align}
\mathcal{L}_{(0,2)}=&\pi\xi\mu g^2\int d^2\theta\,\left( \mathcal{K}_{\bar{z}}  B C+\text{c.c.}  \right)  \\
=&\pi\xi g^2\mu\int d^2\theta\, \log\left( \frac{L_{IR}^2}{ A^\dagger A}\right) \left( A^\dagger B C+\text{c.c.}  \right) \,.
\end{align}
This addition to the Lagrangian does indeed produce the term we suggest in Eq.(\ref{suggest})
\begin{equation}
	 g^2 \mu \chi_R \zeta_R \partial_L \bar{\rho} +\text{ H.c.}
\end{equation}

along with further quartic fermion couplings from the F-terms present in the fermionic multiplets. It is clearly a violation of $\mathcal{N}=(2,2)$ supersymmetry as it involves a fermionic multiplet which does not have a paired bosonic multiplet. 

In total, and once the rescaling of the kinetic logarithms has been performed, the Lagrangian we obtain out of superspace as a result takes the following form:

\begin{align}
\mathcal{L}=& \partial^\mu \bar{\rho} \partial_\mu \rho + i \bar{\chi}\slashed{\partial}\chi + i \zeta^\dagger_R \partial_L \zeta_R +  g^2\mu\left( \zeta_R \chi_R \partial_L \bar{\rho} +\text{ H.c.}\right)\nonumber\\
& +  g^4 \mu^2 \left(\zeta^\dagger_R \zeta_R \right)\left( \chi_L^\dagger \chi_L \right) +  g^4\mu^2\left(\chi^\dagger_R \chi_R \right)   \left(\chi_L^\dagger \chi_L \right)  \,.
\end{align}
This is now manifestly $(0,2)$-supersymmetric, as required.

\section{Conclusion}

We have investigated properties of Supersymmetric Non-Linear Sigma Models that arise as the Lagrangian for semi-local strings in SQED. The scalar modulus $\rho$ that these strings are endowed with seems very different from the internal colour modulus of non-Abelian strings, but we have shown they are similar in at least one aspect: a heterotic deformation affects their worldsheets in very similar ways. 

When a mass is turned on for the gauge scalar multiplet in four dimensions, in both cases, a coupling occurs between fermionic degrees of freedom originally defined in different target spaces on the worldsheet. This breaks the full $(2,2)$ supersymmetry in a way that cannot benefit from any accidental  enhancement. For this structural shape, an explicitly $(0,2)$ superspace action can be written, in a way that clearly violates $(2,2)$ supersymmetry in turn. 

It is nevertheless the case that $\rho$ retains some idiosyncratic features: the asymptotic explicit solution of the field equations that exists in this case proves to be a powerful tool to study the properties of semi-local strings. Given that the modulus $|\rho|$ intervenes in every asymptotic spatial profile we wish to write in the theory, the computation to generate the zero modes and worldsheet theory complicates itself quickly, but subsumes to the expected result eventually. We expect it to become even more difficult to perform, if possible at all, for a large-$\mu$ worldsheet. This exercise will be left for future work.

\section*{Acknowledgments}

This work  is supported in part by DOE grant DE-SC0011842. 
The work of A.Y. was  supported by William I. Fine Theoretical Physics Institute  at the  University 
of Minnesota and  by Russian Foundation for Basic Research Grant No. 18-02-00048.

\begin{appendices}
\section{Conventions}
We work in Euclidean space. We pick the following choices for $\sigma$-matrices
\begin{equation}
\sigma^{\mu\alpha\beta}=\left(\mathbbm{1},\left( \begin{array}{cc}
0 & -i \\ 
-i & 0
\end{array} \right)  ,\left( \begin{array}{cc}
0 & -1 \\ 
1 & 0
\end{array} \right) , \left( \begin{array}{cc}
-i & 0 \\ 
0 & i
\end{array}  \right) \right),\quad  \bar{\sigma}^\mu_{\alpha\beta} = \left( \mathbbm{1},-\sigma^{i\alpha\beta}\right) 
\end{equation}
$SU(2)$ indices, either spinorial or from $R$-symmetry, are contracted with the following tensor
\begin{equation}
\varepsilon^{\alpha\beta}=\left(\begin{array}{cc}
0 & -1 \\ 
1 & 0
\end{array}  \right)  = \varepsilon^{\dot{\alpha}\dot{\beta}},\quad  \varepsilon_{\alpha\beta} = \varepsilon_{\dot{\alpha}\dot{\beta}} = - \varepsilon^{\alpha\beta}
\end{equation}
From our choices in spacetime, the worldsheet gamma matrices necessarily become
\begin{equation}
\gamma=\left( \left(\begin{array}{cc}
0 & 1 \\ 
1 & 0
\end{array}  \right) ,\left(\begin{array}{cc}
0 & i \\ 
-i & 0
\end{array}  \right) \right) 
\end{equation}

\section{Asymptotic Expansions on the Worldsheet}
\label{ap2}

In Section \ref{sec2}, we compute corrections to coefficients of worldsheet couplings. This is an unobvious process conceptually, namely because the field $\chi_R$ is only logarithmically normalisable, though arguments have been put forward that this apparent divergence can be removed safely through field redefinitions. At order $\mu$, for reasons explained above, there is no contribution to the normalisation. One expects them to arrive at higher order in (even) powers of $\mu$. However, we are performing perturbation theory in a setting with an explicit IR cutoff i.e. a maximally large but finite length scale $L$ in the problem. Since $\mu$ has dimensions of mass, one expects that terms dependent on enough powers of $\mu$ will come multiplied by some positive powers of $L$, generically.

In particular, $\lambda_-$ is constructed from a square-log-divergent profile times a factor of $r$, so decays even slower at infinity than $\lambda_+$, thus will lead to a correction that goes as $\mu^2 L^2$. This is symptomatic of doing perturbation theory in settings with IR cutoffs: the full series cannot be trusted, since $\mu$ cannot be smoothly turned off without passing by the IR cutoff regime. This phenomenon is referred to as singular perturbation theory, characteristic of dynamics on multiple scales. The approach is broadly contained in asymptotic analysis, rather than perturbation theory. Hence, we suggest that one should truncate the order at which our series is meaningful.

\section{Useful Transverse Integration Identities}

Computing the transverse integrals yielding worldsheet elements, in the case of semilocal strings, can involve a high number of terms and expressions in the integrand, all contributing towards a small class of possible terms allowed by worldsheet symmetries. It is useful to keep at hand a list of frequently-used identities for quick reference. 

Integrals are performed over the plane transverse to the string solution, and systematically involve functions of the radial coordinate only. Where the integrand can be summed over the entire plane and produce a finite result, that result is used, though some may be required to be cut off, for small $r$ at $\rho$, and for large $r$ at $L_{IR}$. 

The general form of the integrands at hand can usually be reduced to the following type of integral:
\begin{equation}
	\int rdr\, \frac{1}{(r^2+\bar{\rho}\rho)^n}=\frac{1}{n-1}\left(\bar{\rho}\rho \right)^{1-n},\quad n>1
\end{equation}
When $n=1$ the integral requires regularisation:
\begin{equation}
\int rdr\, \frac{1}{(r^2+\bar{\rho}\rho)}=\frac{1}{2}\log\left(\frac{L_{IR}^2}{\bar{\rho}\rho} \right) 
\end{equation}
A combination of both of these two integral types produces the characteristic K\"{a}hler metric of the size modulus. In the deformed worldsheet, at small $\mu$ these formul{\ae} are enough to produce the result.

\end{appendices}

\clearpage

\end{document}